\documentclass[]{pasj02}
\usepackage[switch,mathlines]{lineno}
\begin{document} 
\Received{}
\Accepted{}
\title{A new candidate of a cluster of galaxies behind the Galactic plane, AX J145732$-$5901}

\author{
Shigeo \textsc{Yamauchi}\altaffilmark{1, $\ast$} and Masaru \textsc{Ueno}\altaffilmark{2}
}
\altaffiltext{1}{Faculty of Science, Nara Women's University, Kitauoyanishimachi, Nara, Nara 630-8506, Japan}
\email{yamauchi@cc.nara-wu.ac.jp}
\altaffiltext{2}{Tokyo Metropolitan Bokutoh Hospital, 4-23-15 Kotobashi, Sumida-ku, Tokyo 130-8575, Japan}
\maketitle

\begin{abstract}
AX J145732$-$5901 is an unidentified X-ray source discovered in the ASCA Galactic plane survey. 
Its extended nature and heavily absorbed X-ray spectrum suggest that AX J145732$-$5901 is a cluster of galaxies behind the Galactic plane. 
However, due to limited photon statistics, the spectral shape was not well examined.  
Using the results of the Galactic ridge X-ray emission and Cosmic X-ray background studies based on the Suzaku observations, we reanalyzed the ASCA data of AX J145732$-$5901. 
We confirmed that the source is more extended than the point spread function and the angular size is $14'\times10'$. 
The spectrum was heavily absorbed by interstellar matter equivalent to an $N_{\rm H}$ of $\sim10^{23}$ cm$^{-2}$ and the emission line feature was confirmed. 
The spectrum was represented by a thin thermal plasma model with a temperature of 2.6 keV and a redshift of 0.12. 
Assuming the redshift value, the X-ray luminosity is calculated to be $2.6\times10^{44}$ erg s$^{-1}$ in the 1--10 keV energy band. 
The observational results indicate that  AX J145732$-$5901 is a cluster of galaxies behind the Galactic plane. 
\end{abstract}

\section{Introduction}

Since the beginning of X-ray astronomy, survey observations in the X-ray band have carried out 
and many energetic sources have been discovered in the Universe 
(e.g., Forman et al. 1978; Warwick et al. 1981; McHardy et al. 1981; Wood et al. 1984; Warwick et al. 1988; Elvis et al. 1992). 
Further studies for the detected sources have revealed their nature. 
In 1990's, ROSAT carried out an all-sky X-ray survey in the 0.1--2.4 keV energy band with an angular resolution of  $\approx30''$ 
and detected more than 10$^5$ X-ray sources including newly discovered faint sources (Voges et al. 1999; Boller et al. 2016).
Our understanding of X-ray sources has advanced greatly. 
 
Although X-rays have a strong penetrating power, 
soft X-rays below 3 keV are absorbed by the interstellar medium 
near the Galactic plane (e.g., Snowden et al. 1995). 
On the other hand, since above 3 keV, the interstellar medium is transparent even for line of sight through the Galactic plane,  
hard X-ray observations above 3 keV are an important tool to reveal properties of 
X-ray sources located in the direction of the Galactic plane. 
However, 
since 
spatial resolution and sensitivity of hard X-ray instruments were not enough until 1992, 
detection of faint X-ray sources located in the direction of the Galactic plane has been limited. 

ASCA, launched in 1993, was the first satellite having imaging capability with a broad pass band (0.5--10 keV), a large effective area, and 
moderate energy resolution (Tanaka et al. 1994). 
Hence, ASCA provided us with new finding of faint X-ray sources even through dense matter on the Galactic plane. 

To search for faint X-ray sources located on the Galactic plane, 
we conducted hard X-ray survey on the Galactic plane with ASCA 
(ASCA Galactic plane survey, Yamauchi et al. 2002) and found 
many new X-ray sources, most of which were unidentified sources 
(Sugizaki et al. 2001; Sakano et al. 2002). 
 Some of them are extended sources with a power law-like spectrum, 
 which are proposed to be new non-thermal Galactic supernova remnants (e.g., Bamba et al. 2003). 
A follow up observation for AX J185905$+$0333 
 with Suzaku revealed that the source is a new candidate of a cluster of galaxies behind the Galactic plane (Yamauchi et al. 2011). 
To understand the nature of unidentified faint X-ray sources, further analyses were essential.

\begin{figure*}
  \begin{center}
    \FigureFile(17cm,8cm){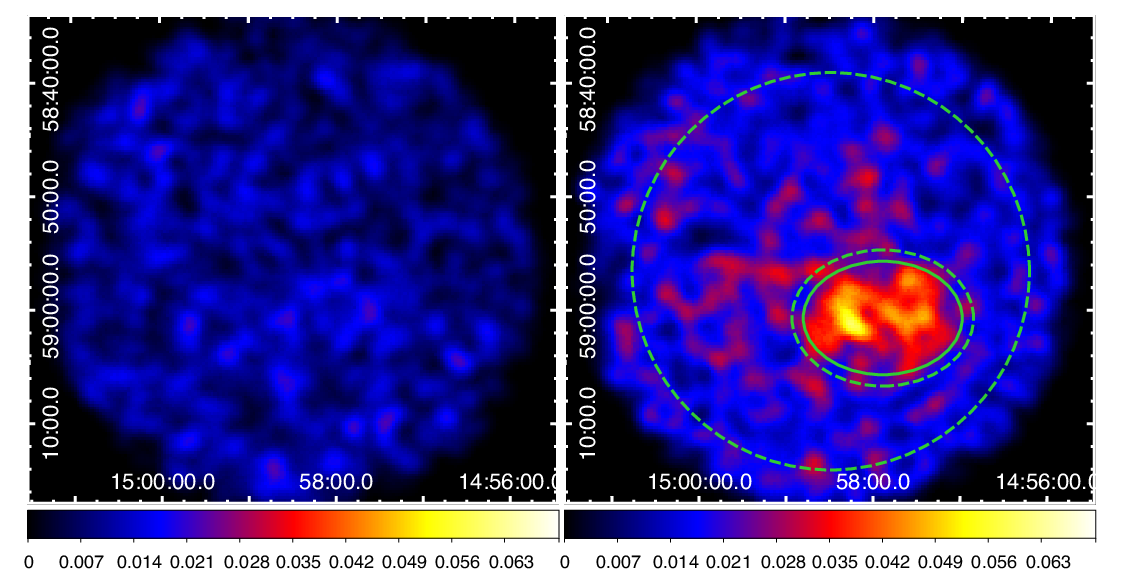}
  \end{center}
  \caption{
X-ray images obtained with the ASCA GIS in the 0.7--2 keV (left) and 2--10 keV (right) energy bands. 
The data of GIS2 and GIS3 were co-added.
The X-ray images were smoothed with a Gaussian distribution  
of $\sigma=3 {\rm pixel}=\sim44''$. 
The coordinates are J2000.0. 
The background subtraction and the vignetting correction are not performed.
The color bar shows X-ray surface brightness in the linear scale (the unit of counts ks$^{-1}$ pixel$^{-1}$).
The green solid and green broken lines show regions from which the source and the background spectra were extracted, respectively. 
}\label{fig:img}
\end{figure*}

AX J145732$-$5901 is one of the unidentified sources 
found in the ASCA Galactic plane survey 
 (Sugizaki et al. 2001).  
Ueno (2005) and Ueno et al. (2006) reported that AX J145732$-$5901 [called G318.6+0.0 in Ueno (2005) and Ueno et al. (2006)] is a heavily absorbed extended source. 
Since an excess emission was seen at 5.9 keV above an absorbed power law continuum with
$\Gamma=2.2^{+1.6}_{-1.1}$ and $N_{\rm H}=(10^{+13}_{-6})\times10^{22}$ cm$^{-2}$, 
AX J145732$-$590 was suggested to be a cluster of galaxies behind the Galactic plane (Ueno 2005). 
However, since the sky background spectrum extracted from nearby sky region was directly subtracted from the source region spectrum, 
each energy bin of the spectrum has a large statistical error, 
and hence a detailed spectral analysis has not been conducted. 

Since AX J145732$-$5901 is located at the low Galactic latitude, 
the most intense sky background is the Galactic ridge X-ray emission (GRXE, Koyama 2018, references therein), 
whose flux is larger than that of the cosmic X-ray background (CXB) in the 1--10 keV band.  
In order to elucidate the spectrum of  AX J145732$-$5901 precisely, we should conduct careful sky background estimation. 
Using Suzaku data, Uchiyama et al. (2013) well modeled the GRXE spectrum with a combination model of several components, whereas  
Kushino et al. (2002) represented the CXB spectrum by a power law function well and studied its intensity variation in the sky. 

Using the results of the GRXE and CXB studies based on the Suzaku observations, we reanalyzed the ASCA data of AX J145732$-$5901.
We firstly modeled the sky background spectrum and then conducted spectral analysis for the source region spectra of AX J145732$-$5901, 
taking account of the contribution of the sky background, instead of the standard method by subtracting the sky background spectrum directly. 
In this paper, we report the results and discuss its nature in detail. 

\section{Data}

The ASCA satellite carried two Solid-state Imaging 
Spectrometers (SIS0, SIS1) and the two Gas Imaging Spectrometers (GIS2, GIS3) placed at the focal plane of the thin foil X-ray Telescope (XRT). 
The Half-Power-Diameter (HPD) of the XRT is about 3 arcmin, whereas 
the field of views of the SIS and the GIS were $22'\times22'$ and  $\sim50'$, respectively. 
The SIS+XRT and the GIS+XRT systems provided us with imaging and spectroscopic observations in the 0.5--10 keV energy band. 
Details of ASCA and the instruments are given in separate papers 
(ASCA satellite: Tanaka et al. 1994; XRT: Serlemitsos et al. 1995; 
SIS: Burke et al. 1991; GIS: Makishima et al. 1996, Ohashi et al. 1996). 

The ASCA Galactic Plane Survey consisted of more than 100 pointing observations with a $\sim$10 ks exposure time for each, 
covering the area of the Galactic inner disk of  $-45^{\circ} < l   < 45 ^{\circ}$ (Sugizaki et al. 2001; Sakano et al. 2002; Yamauchi et al. 2002).
The field centered on (RA, Dec)$_{\rm J2000.0}$=(\timeform{224.D63}, \timeform{-58.D93}) 
[($l$, $b$)=(\timeform{318.D70}, \timeform{-0.D0})] was observed on 1999 Februay 23--24 (Obs. ID 57004070). 
Each SIS camera, consisting of 4-CCD chips, was operated in 2-CCD Bright mode, whereas the GIS was operated in PH mode. 

The ASCA data after screening with the standard criteria were retrieved from the DARTS system operated by ISAS/JAXA.
Since AX J145732$-$5901 has a hard spectrum and was covered by only 1 CCD chip 
of the SIS, 
we analyzed GIS data only. 
The data obtained at the South Atlantic Anomaly, during the earth occultation, at the low elevation angle from the earth rim of $<$5$^{\circ}$, 
and in the high background regions such as at low geomagnetic cut off rigidity were excluded by the standard screening process. 
To reject particle events, the rise-time discrimination technique was also applied. 
The exposure time for GIS2 and GIS3 was 11.2 ks.

Data reduction and analysis were made using the HEAsoft version 6.30.1,  
XSPEC version 12.12.1c, and ds9 version 8.5. 
The abundance tables, the atomic data of the lines and continua of the thin thermal plasma, and the cross-sections of the photoelectric
absorption were taken from Anders \& Grevesse (1989), ATOMDB 3.0.9, and Verner et al. (1996), respectively.

\begin{figure*}
  \begin{center}
    \FigureFile(8cm,8cm){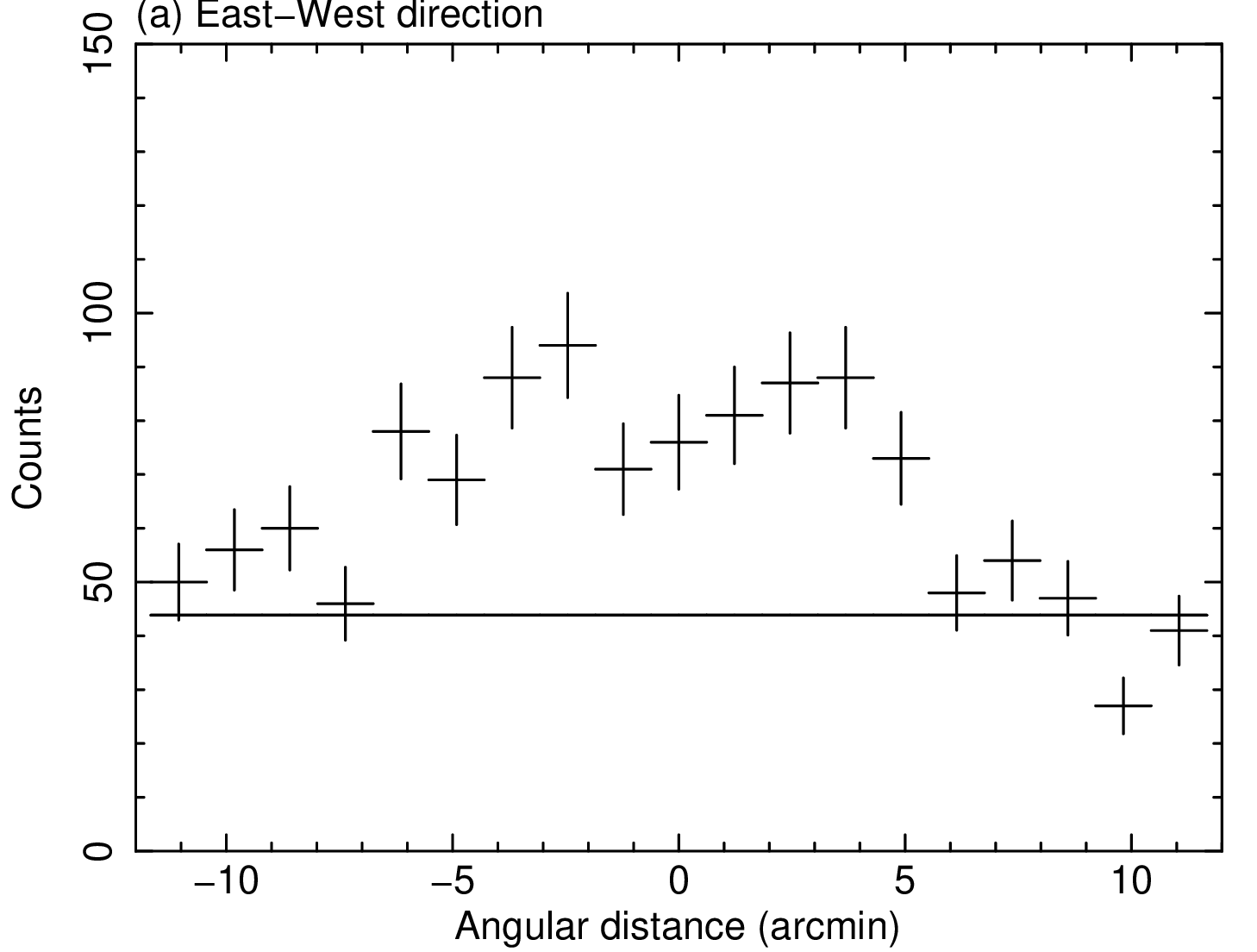}
    \FigureFile(8cm,8cm){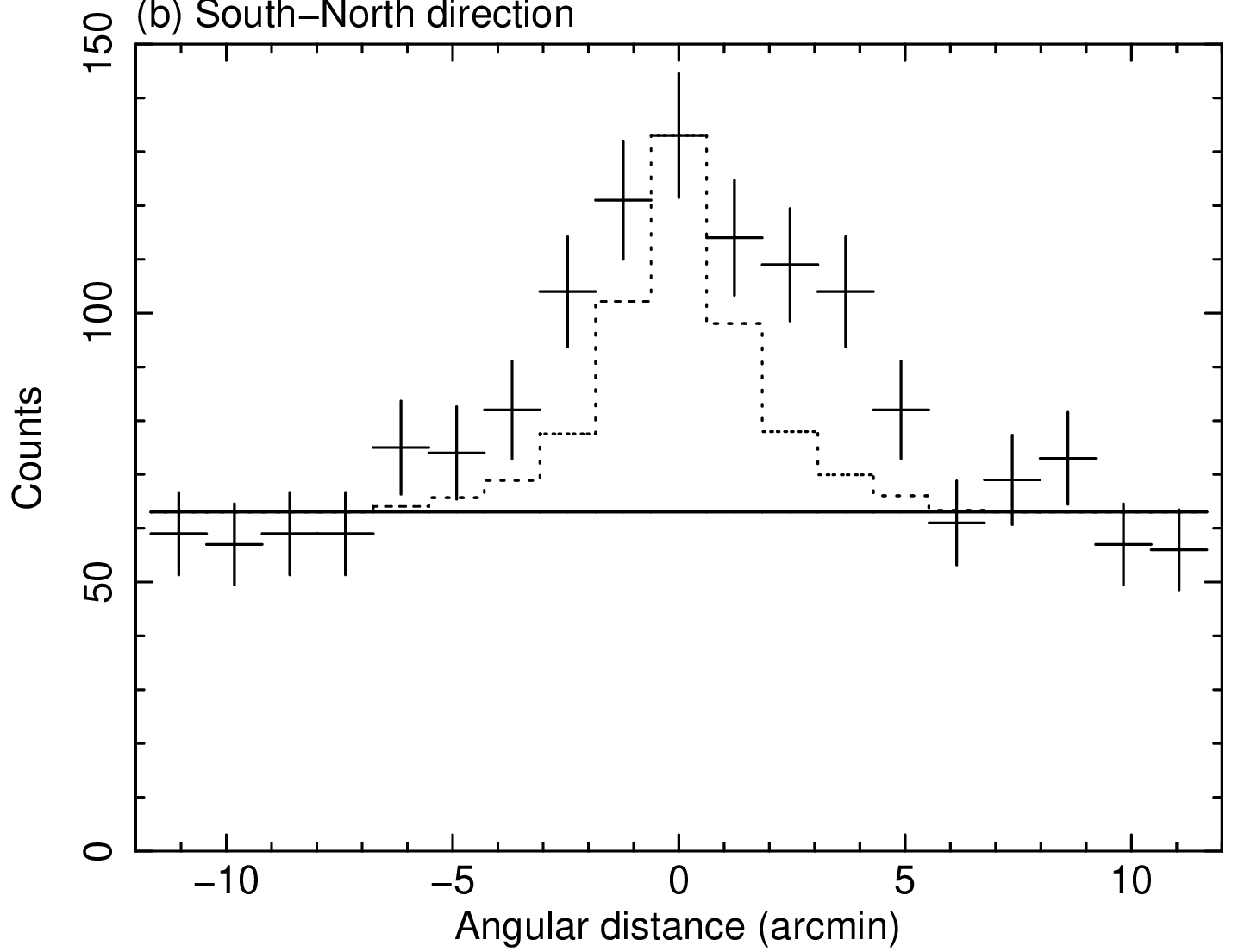}
  \end{center}
  \caption{
Projection profiles of the extended source along (a) a major (East-West) and (b) a minor (South-North) axes. 
The data in the 2--10 keV energy band were used. 
The horizontal axis shows an angular distance from the center of the source, whereas the vertical axis shows an X-ray intensity. 
The vertical errors show 1$\sigma$ level. 
The horizontal solid lines in (a) and (b) show the background level, whereas the dotted line in (b) shows the point spread function of the XRT 
evaluated by using the observational data of 3C 273 (Obs. ID 10701000). 
}\label{fig:proj}
\end{figure*}

\section{Analysis and results}

Figure \ref{fig:img} shows X-ray images in the 0.7--2 keV and 2--10 keV energy bands. 
In the low 
energy band, we found no X-ray source, but in the high energy band, 
we clearly found an extended source at south-west of the center, which is cataloged as AX J145732$-$5901 in Sugizaki et al. (2001). 
We confirmed that AX J145732$-$5901 is a heavily absorbed source  
with an extension of $14'\times 10'$.

\begin{table}[t]
\caption{Best-fitting parameters for the sky background spectrum.}
\begin{center} \label{tab:bg}
\begin{tabular}{lcc} \hline  
	BGD model     		& Parameter               				& Value   \\  \hline 
	FG              		& $N_{\rm H}$ ($10^{22}$ cm$^{-2}$) 	&  0.56$^{\ast}$  \\
					& $kT_{\rm e, FG1}$ (keV)  			& 0.05$^{\ast}$\\
		              		& Norm$_{\rm FG1}^{\dag}$  ($10^{-2}$) 	& 3.4$\pm$1.6  \\
					& $kT_{\rm e, FG2}$ (keV)  			& 0.59$^{\ast}$\\
	         			& Norm$_{\rm FG2}^{\dag}$ ($10^{-5}$) 	& $<$3.2  \\         
	GRXE			& $N_{\rm H}$ ($10^{22}$ cm$^{-2}$) 	&  0.67$^{+1.07}_{-0.23}$ \\
	GRXE$_{\rm LP}$   	& $kT_{\rm e, LP}$ (keV)                		& 1.33$^{\ast}$   \\
	                         		& Norm$_{\rm LP}^{\dag}$ ($10^{-6}$)  	& 2.5$^{+3.5}_{-2.4}$  \\
	GRXE$_{\rm HP}$    & $kT_{\rm e, HP}$ (keV)                		& 6.63$^{\ast}$  \\
	                      		& Norm$_{\rm HP}^{\dag}$ ($10^{-5}$)  	& 1.0$\pm$0.1  \\
	CXB       			& photon index            				& 1.41$^{\ast}$  \\
	                      		& Norm$_{\rm CXB}^{\ddag}$ ($10^{-7}$) & 8.17$^{\ast}$  \\  \hline 
	         			&  $\chi^2$/d.o.f. 					& 109.3/114  \\ 
\hline
\end{tabular}
\end{center}
Notes: The errors show 90\% confidence level. \\
$^{\ast}$ Fixed to the value.\\
$^{\dag}$ Defined as 10$^{-14}$ $\times$$\int n_{\rm H} n_{\rm e} dV$ / (4$\pi D^2$$\Omega$),
where $D$ is the distance (cm), 
$n_{\rm H}$ is the hydrogen density (cm$^{-3}$), 
$n_{\rm e}$ is the electron density (cm$^{-3}$), 
$V$ is the volume (cm$^3$), 
and $\Omega$ is solid angle (arcmin$^{2}$). \\
$^{\ddag}$  In units of photons s$^{-1}$ cm$^{-2}$ arcmin$^{-2}$ at 1 keV. \\
\end{table} 

\begin{figure}
  \begin{center}
    \FigureFile(8cm,8cm){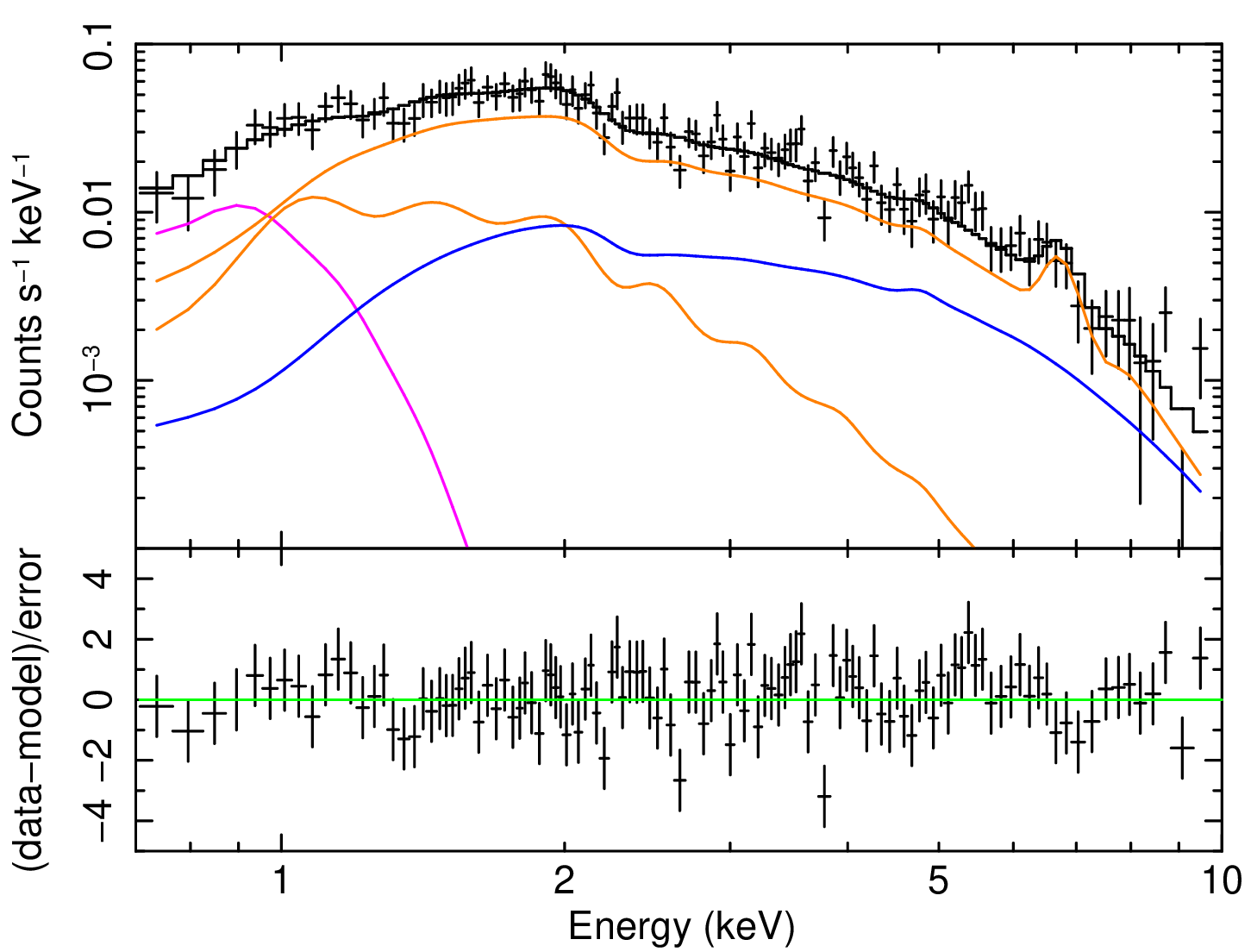}
  \end{center}
  \caption{
The sky background spectrum in the 0.7--10 keV band (upper panel) and residuals from the best-fitting model (lower panel). 
The crosses and histograms in the upper panel show the GIS2$+$GIS3 data and the best-fitting model, respectively. 
The vertical errors show 1$\sigma$ level. 
The FG, GRXE, and CXB components are indicated by the magenta, orange, and blue lines, respectively. 
}\label{fig:bgdspc}
\end{figure}

Figure \ref{fig:proj} displays projection profiles of the extended source along a major (East-West) and a minor (South-North) axes.  
The profiles clearly show that the emission extends more than the point spread function of the XRT. 
On the other hand, the profile along major axis suggests that the source has two bright regions as is seen in figure \ref{fig:img}. 

The X-ray spectrum of AX J145732$-$5901 was extracted from an elliptical region with a major and a minor radii of $7'$ and $5'$, respectively (source region), 
shown by a sold line in figure \ref{fig:img}, whereas 
the sky background data were taken from the outer region in the same FOV, shown by a broken line in figure \ref{fig:img}. 
The non X-ray background (NXB) was estimated from the night-earth data provided by the GIS team. 

First, we fitted the sky background spectrum with an empirical model. 
The observed counts in the 0.7--10 keV energy band were 2013 and 2102 for GIS2 and GIS3, respectively. 
In order to increase photon statistics, we merged the spectra obtained with the GIS2 and GIS3 and grouped the data with a minimum of 30 counts per bin. 
Figure \ref{fig:bgdspc} shows the sky background spectrum after subtracting the NXB. 
We applied a model consisting of the GRXE and the CXB. 
The GRXE spectrum was modeled by a combination of several components, foreground emission (FG) consisting of two thermal components, low-temperature (LP), and 
a high-temperature components (HP), and a cold matter component (CM) consisting of an emission line at 6.4 keV and a non-thermal component (Uchiyama et al. 2013).  
Since no obvious line feature was found at 6.4 keV, we did not include the CM component. 
The temperature of FG, LP, and HP and metal abundances were fixed to those of Uchiyama et al. (2013), 
whereas the parameters of the CXB were fixed to those in Kushino et al. (2002). 
In making the Ancillary Response File (arf), we assumed a uniform sky. 
The best-fitting parameters are listed in table \ref{tab:bg}, and the best-fitting model is plotted in figure \ref{fig:bgdspc}. 

\begin{table}[t]
  \caption{The best-fitting parameters for the AX J145735$-$5901 spectra.}
 \begin{center}  \label{tab:src}
  \begin{tabular}{lc}
   \hline
Parameter& Value\\
\hline 
\multicolumn{2}{c}{Model: (Power law$+$emission line)$\times$absorption}\\
$N_{\rm H}$ ($10^{22}$ cm$^{-2}$)   				&  15$^{+15}_{-8}$ \\
$\Gamma$                                                          		& 3.7$^{+2.8}_{-1.7}$ \\
Normalization$^{\dag}$ ($10^{-2}$)					& 1.9$^{+382}_{-1.8}$\\
$E_{\rm line}$ (keV)                                         			& 5.94$^{+0.20}_{-0.16}$\\
$I_{\rm line}$ ($10^{-5}$ photons s$^{-1}$ cm$^{-2}$)  	& 4.3$^{+1.9}_{-2.0}$ \\
$\chi ^{2}$ /d.o.f.  								&26.7/31\\
\hline 
\multicolumn{2}{c}{Model: (Bremsstrahlung$+$emission line)$\times$absorption}\\
$N_{\rm H}$ ($10^{22}$ cm$^{-2}$)   				&  12$^{+12}_{-6}$ \\
$kT$ (keV)                                                          			& 2.6$^{+5.6}_{-1.5}$ \\
Normalization$^{\ddag}$ ($10^{-3}$)					& 2.6$^{+29.2}_{-2.0}$ \\
$E_{\rm line}$ (keV)                                         			& 5.94$^{+0.20}_{-0.16}$\\
$I_{\rm line}$ ($10^{-5}$ photons s$^{-1}$ cm$^{-2}$)  	& 4.0$^{+1.7}_{-1.8}$ \\
$\chi ^{2}$ /d.o.f.  								&26.4/31\\
\hline 
\multicolumn{2}{c}{Model: APEC$\times$absorption}\\
$N_{\rm H}$ ($10^{22}$ cm$^{-2}$)    				& 13$^{+15}_{-7}$\\
$kT$ (keV)                                                        			& 2.6$^{+4.9}_{-1.5}$ \\
Abundance$^{\ast}$                                     			& $>$0.36\\
Redshift                                                        			& 0.12$^{+0.04}_{-0.03}$\\
Normalization$^{\S}$ ($10^{-3}$)					& 8$^{+69}_{-7}$\\
$\chi ^{2}$ /d.o.f.								& 27.5/31\\
\hline
\end{tabular}
\end{center}
Notes: The errors show 90\% confidence level. \\
$^{\ast}$ Relative to the solar value of Anders \& Grevesse (1989). \\
$^{\dag}$ In units of photons s$^{-1}$ cm$^{-2}$ arcmin$^{-2}$ at 1 keV. \\
$^{\ddag}$ Defined as 3.02$\times$10$^{-15}$$\times$$\int n_{\rm H} n_{\rm e} dV$/(4$\pi D^2$), where $D$ is the distance (cm), $n_{\rm H}$ is the hydrogen density (cm$^{-3}$), $n_{\rm e}$ is the electron density (cm$^{-3}$), and $V$ is the volume (cm$^3$). The unit is cm$^{-5}$.\\
$^{\S}$ Defined as 10$^{-14}$$\times$$\int n_{\rm H} n_{\rm e} dV$/\{4$\pi [D_{\rm A}(1+z)]^2$\}, where $D_{\rm A}$ is the angular distance (cm) 
and $z$ is the redshift. The unit is cm$^{-5}$.\\
\end{table}

\begin{figure}[t]
  \begin{center}
    \FigureFile(8cm,8cm){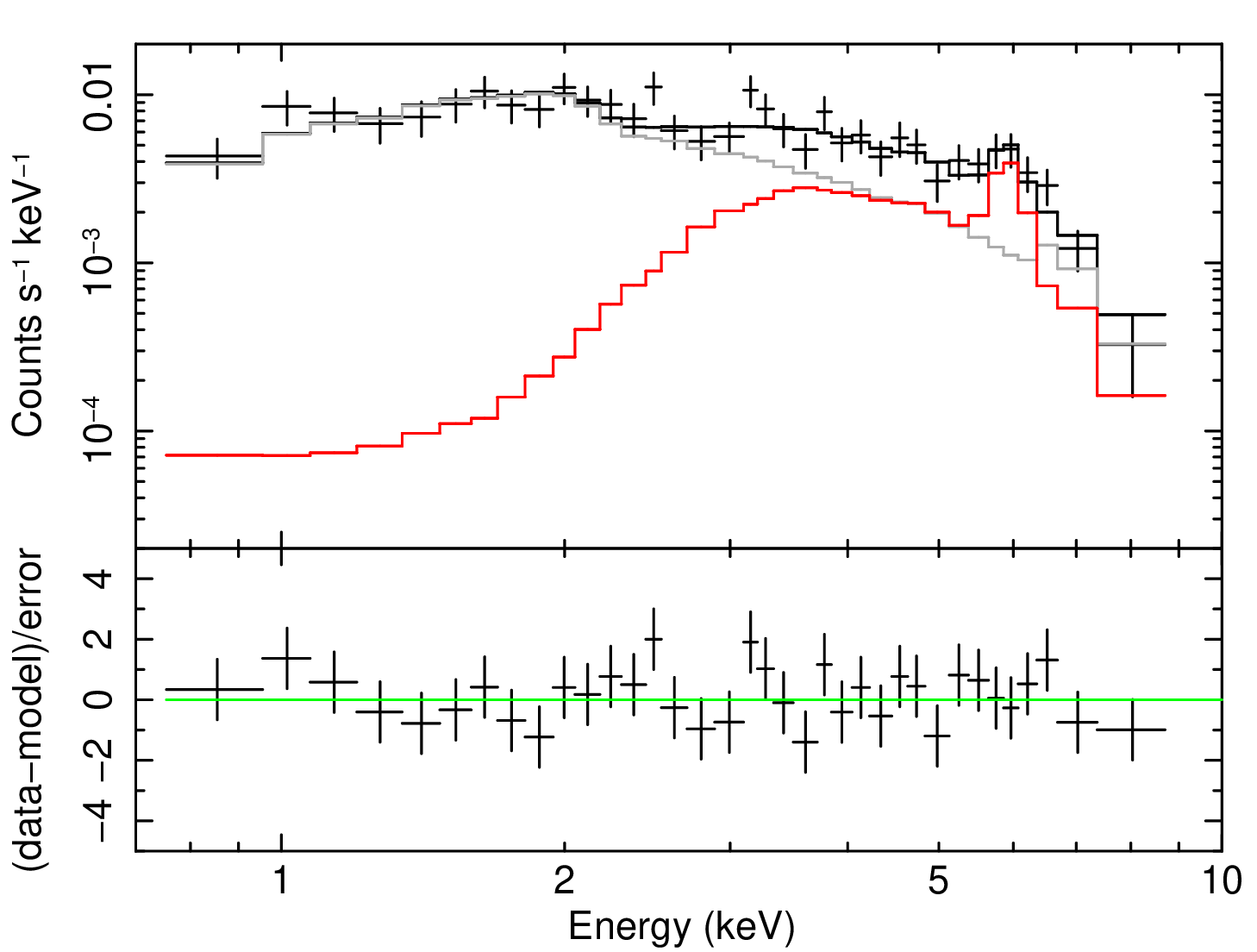}
  \end{center}
  \caption{
X-ray spectrum of AX J145732$-$5901 in the 0.7--10 keV band (upper panel) and residuals from the best-fitting model (lower panel).
The crosses and histograms in the upper panel show the GIS2$+$GIS3 data and the best-fitting model, thin thermal plasma model (red) and the sky background model (gray), respectively. 
The vertical errors show 1$\sigma$ level. 
}\label{fig:srcspc}
\end{figure}

Next, we carried out a spectral analysis for the source region spectrum. 
The X-ray counts within the source region in the 0.7--10 keV energy band including the background counts 
were 404 and 541 for GIS2 and GIS3, respectively. 
We note that 74\% of the total counts were the NXB and the sky background. 
In order to increase photon statistics, we merged the spectra obtained with the GIS2 and GIS3 and grouped the data with a minimum of 25 counts per bin. 
Figure \ref{fig:srcspc} shows the source region spectra after subtracting the NXB. 
The best-fitting sky background model is plotted in the source region spectrum (gray line). 
The figure shows that the intensity of X-ray emission below 2 keV is comparable to that of sky background and excess emission from the sky background is clearly seen above 2 keV, 
which is consistent with that found in the X-ray image (figure \ref{fig:img}). 
 
We fitted the spectrum with a power law function ($\chi^2$/d.o.f.=40.3/33=1.22) and found an emission line feature at $\sim$6 keV in the residuals. 
Therefore, we added a gaussian line model. 
The intrinsic line width was assumed to be null. 
The fit was improved and was accepted ($\chi^2$/d.o.f.=26.7/31=0.86). 
The $\Delta {\chi}^2$ value was 13.6, which shows the F-test probability of 0.002. 
The best-fitting parameters are listed in table \ref{tab:src}. 
The best-fitting line energy is 5.94$^{+0.20}_{-0.16}$ keV. 
The significance level estimated from the obtained photon flux and its statistical error is 3.3$\sigma$.  
We also fitted the spectrum with a bremsstrahlung and an emission line model ($\chi^2$/d.o.f.=26.4/31=0.85). 
The best-fitting parameters are also listed in table \ref{tab:src}. 

We next applied a thermal plasma model ({\tt apec} model in XSPEC). 
Since no atomic line of the abundant element has the energy in the rest frame, 
the line is most likely to be a redshifted iron line. 
To adjust the line structure, we set the redshift parameter free. 
This model also gave an acceptable fit. 
The best-fitting parameters are listed in table \ref{tab:src}, and the best-fitting model is plotted in figure \ref{fig:srcspc}. 
The observed and the $N_{\rm H}$-corrected flux in the 1--10 keV band were calculated to be 
$1.4\times10^{-12}$ erg s$^{-1}$ cm$^{-2}$ and $6.9\times10^{-12}$ erg s$^{-1}$ cm$^{-2}$, respectively. 

Taking account that the intensity variation of the sky background (GRXE+CXB),  
we refitted the spectrum with the sky background by $\pm$10\%, and found that the results are the same within the errors. 

We also examined intensity variation during the observation. 
However, we found no significant variation in the light curve. 

\section{Discussion and conclusion}

Using the results of the GRXE and CXB studies based on the Suzaku observations, we reanalyzed the ASCA data of an unidentified extended source, AX J145732$-$5901. 
The X-ray emission is extended with a size of $14'\times10'$. 
The spectrum was heavily absorbed and exhibited an emission line at 5.94 keV. 
Using the method of Willingale et al. (2013)\footnote{https://www.swift.ac.uk/analysis/nhtot/index.php}, 
the Galactic $N_{\rm H}$ value toward the line of sight of AX J145732$-$5901 is calculated to be 1.9$\times$10$^{22}$ cm$^{-2}$.   
The observed value of $\sim$10$^{23}$ cm$^{-2}$ is larger than the calculated value in the line of sight.  
Thus, the heavily absorbed X-ray spectrum suggests that AX J145732$-$5901 is likely to be an extragalactic source. 

The extended nature, the temperature of a thin thermal plasma model of 2.6 keV, and the Fe abundance of $>$0.36 solar are well
consistent with those of clusters of galaxies (e.g., Fukazawa et al. 2004). 
 Assuming the redshift of 0.12, the Hubble constant of H$_0$=70 km s$^{-1}$, $\Omega_{\rm M}$=0.3, and  $\Omega_{\Lambda}$=0.7, 
 the luminosity distance and angular distance are estimated to be 560 Mpc and 440 Mpc, respectively. 
The X-ray luminosity is calculated to be $2.6\times10^{44}$ erg s$^{-1}$ in the 1--10 keV energy band.
Comparing the relation between the temperatures and luminosities derived from a large sample of clusters of galaxies 
(e.g., Arnaud \& Evrard 1999; Xue \& Wu 2000; Fukazawa et al. 2004; Pratt et al. 2009),  
we found that the observed temperature and luminosity are consistent with the relation. 

Using the best-fitting normalization of the apec model fit of $8\times10^{-3}$ cm$^{-5}$, 
we can calculate the volume emission measure to be $2.4\times10^{67}$ cm$^{-3}$. 
Assuming the volume to be ellipsoidal with a major radius $r_1$ of 7$'$ and a minor radius $r_2$ of 5$'$, 
we calculate the volume $V=\frac{4\pi}{3}r_1 r_2^2 \sim 4.7\times10^{73}$cm$^{3}$. 
Then, we obtain the density of $\sim7\times10^{-4}$ cm$^{-3}$ and the total gas mass of $\sim3\times10^{13}$ M$_{\odot}$. 
If the gas fraction is 15\% of the total mass of clusters of galaxies, 
the total mass is estimated to be $\sim2\times10^{14}$ M$_{\odot}$. 

The angular size of $14'\times10'$ corresponds to 1.8 Mpc$\times$1.3 Mpc. 
The emission is elongated in the East-West direction and seems to have local structures. 
The profile of the South-North direction (figure \ref{fig:proj}b) is roughly approximated by the isothermal $\beta$ model (source)$+$constant model (background) 
with $\beta$ value of $\sim0.5$ and a core radius of $\sim1'-2'$ corresponding to $\sim120-250$ kpc.

All the observational results support the idea that  AX J145732$-$5901 is a cluster of galaxies behind the Galactic plane. 
The obtained X-ray morphology may suggest an unrelaxed (merging) cluster of galaxies. 
However, due to limited photon statistics in short exposure time observation, we cannot examine the surface brightness distribution in detail. 
In addition, the spectral parameters derived from the present observation have still large uncertainties. 
To clarify the characteristics, further observations with good photon statistics are required. 

We note that the observed $N_{\rm H}$ value of $\sim$10$^{23}$ cm$^{-2}$ is several times 
larger than the Galactic $N_{\rm H}$ value calculated using the method of Willingale et al. (2013) as is described above. 
Since clusters of galaxies have no intrinsic absorption associated with the source, 
this difference suggests that the calculation from the method underestimates the $N_{\rm H}$ value in the case close to the Galactic plane. 

Clusters of galaxies have evolved throughout the history of the universe and are a key component of a large scale structure in the Universe, 
and hence finding clusters of galaxies is important to understand evolution of the Universe. 
However, 
observations in the optical band is hard to find objects behind the Galactic plane due to obscuring by a large amount of interstellar matter. 
Therefore, only a small fraction of  clusters of galaxies has been found near to the Galactic plane, so called zone of avoidance. 
With XMM-Newton, several candidates have been discovered in the zone of avoidance (e.g., Nevalainen et al. 2001; Lopes de Oliveira et al. 2006; Barri\`{e}re  et al. 2015). 
Suzaku observations on the Galactic plane also have discovered candidates of clusters of galaxies (Yamauchi et al. 2010; Yamauchi et al. 2011; Mori et al. 2013; Nobukawa et al. 2015)
and here we reported a new candidate. 
An X-ray observation is powerful tool to find clusters of galaxies hidden by interstellar matter in the Galaxy. 
To reveal the large scale structure behind the Galaxy, X-ray survey observations are encouraged. 

\ack
This work was supported by the Japan Society for the Promotion of Science (JSPS) KAKENHI Grant Numbers 21K03615 and 24K00677 (SY).


\end{document}